\documentclass[twocolumn]{aastex631}
\usepackage{amsmath}
\usepackage{CJK}

\newcommand\cloudy{\texttt{CLOUDY}}
\newcommand\cue{\texttt{Cue}}
\newcommand\QH{Q_\mathrm{H}}
\newcommand\Qion{Q_\mathrm{ion}}
\newcommand\QHeII{Q_\mathrm{HeII}}
\newcommand\rHeII{\frac{Q_\mathrm{HeII}}{Q_\mathrm{H}}}
\newcommand\nH{n_\mathrm{H}}
\newcommand{\Fuv}{F_\mathrm{1500}}
\newcommand\eff{\xi_\mathrm{ion}}
\newcommand\fstar{f_\mathrm{star}}
\newcommand\bL{\boldsymbol{L}}

\newcommand\bsigma{\boldsymbol{\sigma}}

\begin{document}
\begin{CJK*}{UTF8}{gbsn}

\title{No top-heavy stellar initial mass function needed: the ionizing radiation of GS9422 can be powered by a mixture of AGN and stars}

\author[0000-0002-0682-3310]{Yijia Li (李轶佳)}
\email{yzl466@psu.edu}
\affiliation{Department of Astronomy \& Astrophysics, The Pennsylvania State University, University Park, PA 16802, USA}
\affiliation{Institute for Gravitation and the Cosmos, The Pennsylvania State University, University Park, PA 16802, USA}

\author[0000-0001-6755-1315]{Joel Leja}
\affiliation{Department of Astronomy \& Astrophysics, The Pennsylvania State University, University Park, PA 16802, USA}
\affiliation{Institute for Computational \& Data Sciences, The Pennsylvania State University, University Park, PA 16802, USA}
\affiliation{Institute for Gravitation and the Cosmos, The Pennsylvania State University, University Park, PA 16802, USA}

\author[0000-0002-9280-7594]{Benjamin D. Johnson}
\affiliation{Center for Astrophysics $\mid$ Harvard \& Smithsonian, Cambridge, MA, USA}

\author[0000-0002-8224-4505]{Sandro Tacchella}
\affiliation{Kavli Institute for Cosmology, University of Cambridge, Cambridge, UK}
\affiliation{Cavendish Laboratory, University of Cambridge, Cambridge, UK}

\author[0000-0003-2895-6218]{Rohan P.\ Naidu}
\affiliation{MIT Kavli Institute for Astrophysics and Space Research, 77 Massachusetts Ave., Cambridge, MA 02139, USA}
\altaffiliation{NHFP Hubble Fellow}

\begin{abstract}
JWST is producing high-quality rest-frame optical and UV spectra of faint galaxies at $z>4$ for the first time, challenging models of galaxy and stellar populations. One galaxy recently observed at $z=5.943$, GS9422, has nebular lines and UV continuum emission that appears to require a high ionizing photon production efficiency. This has been explained with an exotic stellar initial mass function (IMF), 10-30x more top-heavy than a Salpeter IMF \citep{Cameron2024}. Here we suggest an alternate explanation to this exotic IMF. We use a new flexible neural net emulator for CLOUDY, \cue{}, to infer the shape of the ionizing spectrum directly from the observed emission line fluxes. By describing the ionizing spectrum with a piece-wise power-law, \cue\ is agnostic to the source of the ionizing photons. \cue\ finds that the ionizing radiation from GS9422 can be approximated by a double power law characterized by $\rHeII = -1.5$, which can be interpreted as a combination of young, metal-poor stars and a low-luminosity active galactic nucleus (AGN) with $F_{\nu} \propto \lambda ^ {2}$ in a 65\%/35\% ratio. This suggests a significantly lower nebular continuum contribution to the observed UV flux (24\%) than a top-heavy IMF ($\gtrsim80$\%), and hence, necessitates a damped Lyman-$\alpha$ absorber (DLA) to explain the continuum turnover bluewards of $\sim1400$\,\AA. While current data cannot rule out either scenario, given the immense impact the proposed top-heavy IMF would have on models of galaxy formation, it is important to propose viable alternative explanations and to further investigate the nature of peculiar high-z nebular emitters.
\end{abstract}

\section{Introduction}
The stellar initial mass function (IMF) is the mass distribution of zero-age main sequence stars, and is a critical ingredient in a wide variety of astrophysical applications, including dynamical studies, metal production rates, supernova and gravitational wave events, black hole seed production, and galaxy spectral energy distribution (SED) modeling. Potential changes in the shape of the IMF, particularly the slope or the cutoff of the massive part of the IMF, are a dominant systematic effect in many of these fields; for example, reasonable bracketing choices for the IMF can alter the inferred mass and star formation rate of high-redshift galaxies by a factor of 10, a systematic uncertainty some 2-5x larger than standard measurement uncertainties (e.g., \citealt{Woodrum2023}; \citealt{Wang2024}). Despite its importance, there is no consensus on how or whether the IMF varies as a function of cosmic time, and metallicity, or other variables as direct constraints are challenging due to a combination of the degeneracy between standard galaxy parameters and the IMF and the intrinsic faintness of main-sequence stars in integrated light (e.g., \citealt{Conroy2013}). A number of studies have constrained the IMF in small samples of galaxies using dynamical modeling, lensing, or deep spectra (see \citealt{Hopkins2018} and references therein), but these works require very high S/N data or specific lensing geometries which, to date, have made it impossible to conduct detailed large-scale study. The issue of the IMF has become particularly relevant given JWST's observations of extremely luminous galaxies in the early universe (e.g., \citealt{Haslbauer2022}; \citealt{Naidu2022}; \citealt{Bunker2023a}; \citealt{Labbe2023};
\citealt{Tacchella2023a}); while many possible explanations have been proposed, one plausible solution is a cosmological evolution of the IMF, making these galaxies unusually rich in O and B stars (e.g., \citealt{Steinhardt2022}).

One of the most direct observational probes of the number of O and B stars in unresolved stellar populations is nebular emission. Nebular emission is shaped by the chemical abundances, density, ionization states of the gas as well as the ionizing sources themselves (e.g., \citealt{Kewley2019}; \citealt{Cameron2023c}; \citealt{Sanders2023}; \citealt{Hirschmann2023b}), and plays an important role on the interpretation of the photometry and spectra of high-redshift galaxies (e.g., \citealt{Stark2013}; \citealt{Tacchella2023}; \citealt{Trussler2023}). In principle, if stars dominate the ionizing energy, the balance between the observed stellar continuum and the observed nebular emission provides some sensitivity to the IMF (e.g., \citealt{ArrabalHaro2023}; \citealt{Cameron2024}). Intriguingly, recent JWST observations have unveiled individual galaxies with unusual nebular emission properties in the early universe (e.g., \citealt{ArrabalHaro2023}; \citealt{Bunker2023a}; \citealt{Cameron2024}; \citealt{Katz2023}; \citealt{Isobe2023}). These galaxies defy simple interpretation using nebular models calibrated in lower-redshift star-forming galaxies, and in some cases more exotic solutions such as very early AGN 
have been proposed (e.g., GNz-11; \citealt{Oesch2016};
\citealt{Bunker2023a}; 
\citealt{Tacchella2023a};
\citealt{Maiolino2024}).

Recently, \citet{Cameron2024} argued that a galaxy at $z = 5.943$, JADES-GS$+$53.12175$-$27.79763 (hereafter GS9422), shows evidence for a top-heavy IMF. The JWST/NIRSpec observation of GS9422 was acquired by the JWST Advanced Deep Extragalactic Survey (JADES) \citep{Eisenstein2023} in four grating modes, G140M/F070LP, G235/F170LP, G395M/F290LP and G395H/F290LP, with 7 hr integration for each, and in the low-resolution PRISM/CLEAR mode with 28 hr integration. \citet{Cameron2024} highlights two main observational constraints that disfavor typical ionizing sources; a UV turnover at $\lambda \approx 1430$\,\AA\ combined with a Balmer jump suggest strong two-photon nebular continuum emission and high neutral hydrogen absorption for Ly$\alpha$; and the detected HeII emission lines are too faint for AGN, or X-ray binaries (e.g., \citealt{Garofali2024}). A stellar population with an exotic, top-heavy IMF including 10-30x more young massive stars/Wolf-Rayet stars (WR) stars than the typical IMF can solve the above two problems. However, as a top-heavy IMF would represent an extraordinary and unprecedented scenario, it is prudent to explore other possibilities to explain the observed spectral features.

To investigate the intriguing nebular nature of GS9422, we adopt a flexible nebular tool \cue\ \citep{Li2024}. It is a neural network emulator around the photoionization code \cloudy\ (\citealt{Ferland1998}; \citealt{Cloudy2023}), and is built specifically to model the nebular continuum and line emission from different ionizing sources in galaxies. A core feature of \cue\ is that it parameterizes the ionizing spectrum with a flexible four piece-wise power law. In this way, we are agnostic to the actual identity of the ionizing source and able to marginalize over or infer the ionizing spectrum directly. Also, by training a large number of photoionization models into a compact neural network (NN), \cue\ achieves a significant speedup at evaluating the nebular emission over a broad model space. The wide coverage in the ionizing radiation and nebular parameter space makes this tool suitable for addressing challenges posed by the unique chemistry and ionizing properties of galaxies with fast MCMC analysis.

In this paper we perform a source-agnostic exploration of the nebular line and continuum emission in GS9422 in search of explanations outside of an exotic top-heavy IMF. The paper is structured as follows. In Section~\ref{sec:cue_fit}, we model the emission line fluxes measured from JWST NIRSpec spectroscopy \citep{Cameron2024} with \cue\, and infer the ionizing spectrum shape and nebular parameters directly. In Section~\ref{sec:cue_res}, we demonstrate that our model emission lines show a good fit to the observations and discuss the inference results from \cue.  In Section~\ref{sec:ssp_agn}, we show that the observed line emission and 1500\,\AA\ flux density can be explained using a combination of standard stellar populations and a low-luminosity AGN, without any IMF variations. In Section~\ref{sec:summary}, we summarize and discuss observational evidence for and against the proposed scenario.


\section{inferring the ionizing properties from the observed rest-UV/optical spectrum}\label{sec:cue_fit}

\begin{figure*}
    \centering
    \includegraphics[width=0.9\textwidth]{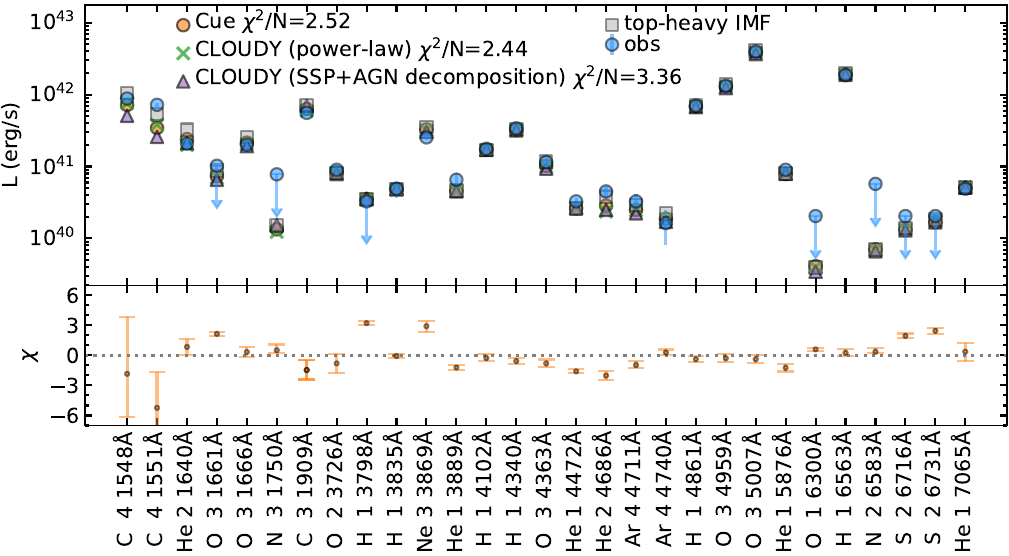}
    \caption{Emission line fit results. Upper panel: emission line measurements along with their uncertainties from \citet{Cameron2024} are in blue. The \cue\ posterior median assuming a double power-law ionizing spectrum and five free nebular parameters is in orange. The nine free parameters of the \cue\ fit are shown in Figure~\ref{fig:cue_corner}. To confirm the \cue\ result, we also run \cloudy\ using the ionizing spectrum and nebular parameters \cue\ predicted as input, and the output line luminosities are shown in green. Additionally \cloudy\ prediction from an SSP ($t_\mathrm{age} = 3.2$\,Myr and $\log (Z / Z_\odot) = -1.34$) with the proposed top-heavy IMF is in grey. The \cloudy\ output given a combination of a MIST+C3K SSP and an AGN ionizing spectrum as the input (see Section~\ref{sec:ssp_agn}) is in purple. Lower panel: the residual of the \cue\ fit results denominated by the measurement uncertainties. The error bars indicate the 1$\sigma$ posterior range of $\chi$. The results of \cue\ fit and the SSP+AGN fit match the observations reasonably well.}
    \label{fig:cue_lines}
\end{figure*}

\begin{figure}
    \centering
    \includegraphics[width=0.45\textwidth]{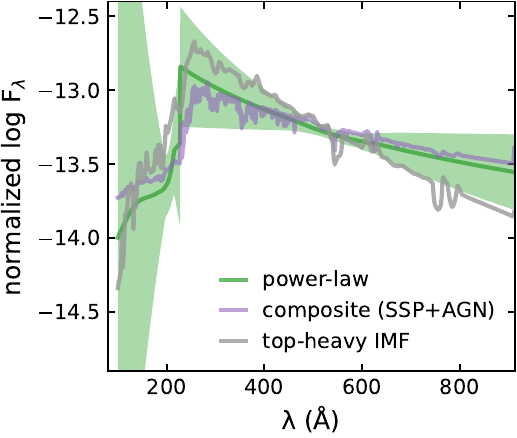}
    \caption{Ionizing spectra corresponding to the models in Figure~\ref{fig:cue_lines}. The green line and the shaded region represent the median and 1$\sigma$ range of the \cue\ inferred ionizing spectrum, respectively. The purple line is the proposed decomposition of the inferred ionizing spectrum from \cue\ into stellar populations and AGN. The grey line is a top-heavy IMF SSP ($t_\mathrm{age} = 3.2$\,Myr and $\log (Z / Z_\odot) = -1.34$) estimated from \citet{Cameron2024} for comparison. The composite scenario matches the \cue\ fit results well, and produces fewer HeII ionizing photons than the top-heavy SSP.}
    \label{fig:ion_spec}
\end{figure}
\citet{Cameron2024} demonstrates the challenges in describing the spectroscopy of GS9422 with a typical ionizing source, such as a simple stellar population (SSP), an AGN, WR stars in a stellar population with a typical IMF, and an X-ray binary. The unknown nature of the ionizing source motivates the use of \cue, a flexible nebular emission emulator for CLOUDY that can fit an exotic source, or a combination of different ionizing sources. Here, we use \cue\ v0.1 to constrain the ionizing spectrum and nebular parameters directly from the observed emission line fluxes. 

\subsection{\cue: a neural-net emulator for CLOUDY}
\cue\ emulates the spectral synthetic code \cloudy\ \citep{Cloudy2023} to calculate the continuum and line emission from a single HII region. It assumes a spherical shell gas cloud geometry and a constant gas density $\nH$ across the region. The distance from the central source to the inner face of the cloud is fixed at $R_\mathrm{inner} = 10^{19}$\,cm. The gas-phase metallicity is specified by [O/H], and \cue\ further allows freedom in [C/O] and [N/O]. Importantly, the ionizing radiation is modeled by a flexible 4-part piecewise-continuous power-law. The edges of the power laws follow the ionization potential of the HeII, OII, and HeI. The permitted slopes of this piece-wise power-law are tuned to ensure that the emulator can describe a wide range of ionizing sources, including stellar models, AGNs, post-AGBs, and a mixture of those. The normalization of the ionizing spectrum is set by the ionizing photon production rate $\QH = 4 \pi R_\mathrm{inner}^2 c \nH U$, where $U$ is the ionization parameter. In total, \cue\ has 5 free nebular parameters, $U$, $\nH$, [O/H], [N/O], and [C/O], and 7 free parameters characterizing the shape of the ionizing spectrum. This flexibility allows \cue\ to model unusual ionizing sources and extreme nebular conditions across different redshifts --- including GS9422. Our emulator consists of 1 NN for nebular continuum and 14 NNs for emission line prediction. It is trained on $2 \times 10^6$ \cloudy\ models of the aforementioned 12 free parameters. \cue\ demonstrates a high accuracy, with $\sim$ 1\% uncertainty in predicting the full 1000\,\AA\--10\,mm nebular continuum and $\sim$ 5\% uncertainty in 128 UV to far-infrared emission lines. Mock tests suggest Cue is well-calibrated and produces useful constraints on the ionizing spectra when $S/N (\mathrm{H}_\alpha) \gtrsim 10$. A full description of \cue\ v0.1 is in \citet{Li2024}.

According to \citet{Cameron2024}, the HeII emission line flux from GS9422 is in particular challenging to explain as it is too faint for an AGN-dominated galaxy, yet too bright to be powered by young stellar populations with a typical IMF. We therefore employ \cue\ to model the ionizing spectrum of this object in the form of two power laws separated by the HeII ionization edge ($\lambda_\mathrm{HeII} = 228\,\mathrm{\AA}$). Although \cue\ permits freedom in the slope and the normalization of four ionizing spectrum segments, we simplify the number of free power laws to two. In this way, we allow the HeII ionizing spectrum to be entirely free while ensuring continuity at the other ionization edges, as expected in standard ionizing spectra. In this model, the ionizing spectrum flux density is given by
\begin{equation}
    F_\nu = 
    \begin{cases}
    A_\mathrm{HeII} \lambda^{\alpha_\mathrm{HeII}} & \mathrm{for}~1\,\mathrm{\AA}<\lambda<228\,\mathrm{\AA};
    \\
    A_\mathrm{HI} \lambda^{\alpha_\mathrm{HI}} & \mathrm{for}~228\,\mathrm{\AA}<\lambda<912\,\mathrm{\AA}.
    \end{cases}
\end{equation}
In the actual \cue\ fits, $A_\mathrm{HeII}$ and $A_\mathrm{HI}$ are transformed from the $\QH$ and the flux ratio of the two power laws $F_\mathrm{HI}/F_\mathrm{HeII} \equiv \frac{\int_{228\mathrm{\AA}}^{912\,\mathrm{\AA}} F_\nu \mathrm{d}\nu}{\int_{1\mathrm{\AA}}^{228\,\mathrm{\AA}} F_\nu \mathrm{d}\nu}$. In this way, \cue\ describes the shape of the ionizing spectrum by three parameters, $\alpha_\mathrm{HeII}$, $\alpha_\mathrm{HI}$, and $F_\mathrm{HI}/F_\mathrm{HeII}$. 

We have introduced five \cue\ parameters describing the nebular conditions and three \cue\ parameters prescribing shape of the ionizing spectrum. 
Because \cue\ is an emulator for a single homogeneous HII region, we scale its predictions by a factor of $\Qion \over \QH$ to model the nebular emission from the entire galaxy as it comes from a cluster of multiple HII regions of the same gas properties. Hence, $\Qion$, the effective ionizing photon rate, becomes the last free parameter of the our nebular model. 
In summary, our emission line fit have 9 free parameters, $\alpha_\mathrm{HeII}$, $\alpha_\mathrm{HI}$, $F_\mathrm{HI}/F_\mathrm{HeII}$, $U$, $\nH$, [O/H], [N/O], [C/O], and $\Qion$.

\subsection{Observations and Sampling}
We sample the parameter posterior distribution using the dynamic nested sampling code \texttt{dynesty} \citep{Speagle2020,Koposov2022}. A uniform prior is adopted for the 9 free parameters (see their allowed range in \citealt{Li2024}). A Gaussian likelihood is assumed.
Our fit data includes 22 lines that have measured fluxes and and uncertainties from \citet{Cameron2024}. Note that we do not include Ly$\alpha$ because its interpretation is subject to complex factors such as the absorption by the neutral intergalactic medium (IGM) and the interstellar medium (ISM), and resonance scattering (e,g., \citealt{Smith2019}). Later we will show that in our preferred model of GS9422, Lyman-$\alpha$ is additionally affected by absorption (see Section~\ref{sec:summary}). 
Apart from the 22 measured lines, we include 7 lines with 3$\sigma$ upper limits detected as well, and fit them assuming zero flux and an associated uncertainty of $\bsigma_{\mathrm{up}} = \frac{1}{3} \bL_{\mathrm{up}}$. When a line has reported measurements or upper limits both in the NIRSpec Prism/CLEAR spectrum or the gratings by \citet{Cameron2024}, we adopt both measurements. There are five additional lines with reported upper limits that \cue\ does not emulate and not used in the fit, including four high ionization state NV 1239\,\AA, NV 1243\,\AA, NIV] 1483\,\AA,  and NIV] 1486\,\AA, and a high-order Balmer line H11 (3771\,\AA). As a final check on the accuracy of the neural net, we run \cloudy\ using the median inferred nebular properties and ionizing spectrum from \cue\ in the next section, and confirm that the \cloudy\ output agree with the upper limits of these lines.

Since the reported Balmer decrements (see Fig. 5 of \citealt{Cameron2024}) suggest effectively zero dust reddening in GS9422, we assume the measurements provided by \citet{Cameron2024} require no dust corrections. 
We also impose an S/N = 10 ceiling on the observed emission line fluxes, accounting for any potential issues in the NIRSpec flux calibration and in differential slit-losses -- see the NIRSpec analyses, e.g., \citealt{Bunker2023}; \citealt{Curtis-Lake2023}; \citealt{Heintz2024}; \citealt{deGraaff2024}. This also helps reconcile any offsets between line fluxes measured from the PRISM and gratings (e.g., \citealt{Bunker2023}). We also take into account the emulator uncertainty by adding it to $\bsigma$ in quadrature, for most lines this is of order $<$5\% (see \citealt{Li2024}). 

\section{Results from inference with \cue}\label{sec:cue_res}

The \cue\ settings described in Section~\ref{sec:cue_fit} are used to fit the emission lines and infer the ionizing spectrum and nebular properties of GS9422.

First, Figure~\ref{fig:cue_lines} shows that \cue\ is able to achieve a good fit to the observations. The median of the emission line posteriors mostly lie within $\lesssim 2\sigma$ of the observations on average with a $\chi^2$ of 2.52 of the median inferred emission lines. This suggests that the double-power-law detailed in Section~\ref{sec:cue_fit} is sufficient to describe the ionizing radiation from this galaxy. Figure~\ref{fig:cue_lines} indicates that most emission lines have their measurement consistent with $1\sigma$ of the posterior. Specifically, the observed HeII 1640\,\AA\ and HeII 4868\,\AA\ fluxes are reproduced within $\lesssim 2\sigma$. This is important in that it shows the double power-law ionizing spectrum can accurately reproduce the relatively small ratio of He$^{+}$ ionizing photons to hydrogen ionizing photons. Notably, \cue\ does not predict additional emission lines bright enough to be observed in the rest-frame wavelength coverage of the PRISM spectrum beyond those already measured in \citet{Cameron2024}. Specifically, \citep{Cameron2024} finds strong MgII doublets at 2796\,\AA\ and 2803\,\AA\ in their model predictions. But with our inferred nebular parameters, \cloudy\ estimates weak MgII doublets for all ionizing spectra shown in Figure~\ref{fig:ion_spec}.

Only two lines, [CIV] 1551\,\AA\ and [NeIII] 3869\,\AA\ demonstrate a $\gtrsim 3\sigma$ deviation between the \cue\ predictions and the detected flux. [CIV] 1551\,\AA\ is reported to be partially blended with [CIV] 1548\,\AA\ in the G140M spectrum \citep{Cameron2024}. Since \cue\ is able to reproduce the observed total flux of the [CIV] doublets within 1$\sigma$, we consider the model [CIV] fluxes to be generally consistent with the observation. Also, the inferred [CIV] lines have large uncertainties as they are sensitive to the detailed ionizing structure of the gas cloud and processed in a non-trivial manner by the ISM. [NeIII] 3869\,\AA\ has two neighboring lines, H10 (3798\,\AA) and HeI 3889\,\AA, and the total flux of these three lines predicted by \cue\ matches the observation within 2$\sigma$.  In addition, since Ne$^{++}$ is iso-electric equivalent to O$^{++}$ and these two ions have similar energy levels, [NeIII] and [OIII] lines respond similarly to the nebular model variations. Hence, the model space allowed for [NeIII] 3869\,\AA\ is constrained by the fits to [OIII] lines. 

\subsection{Inferred Nebular Conditions}\label{subsec:cue_neb}
\begin{figure*}
    \centering
    \includegraphics[width=0.98\textwidth]{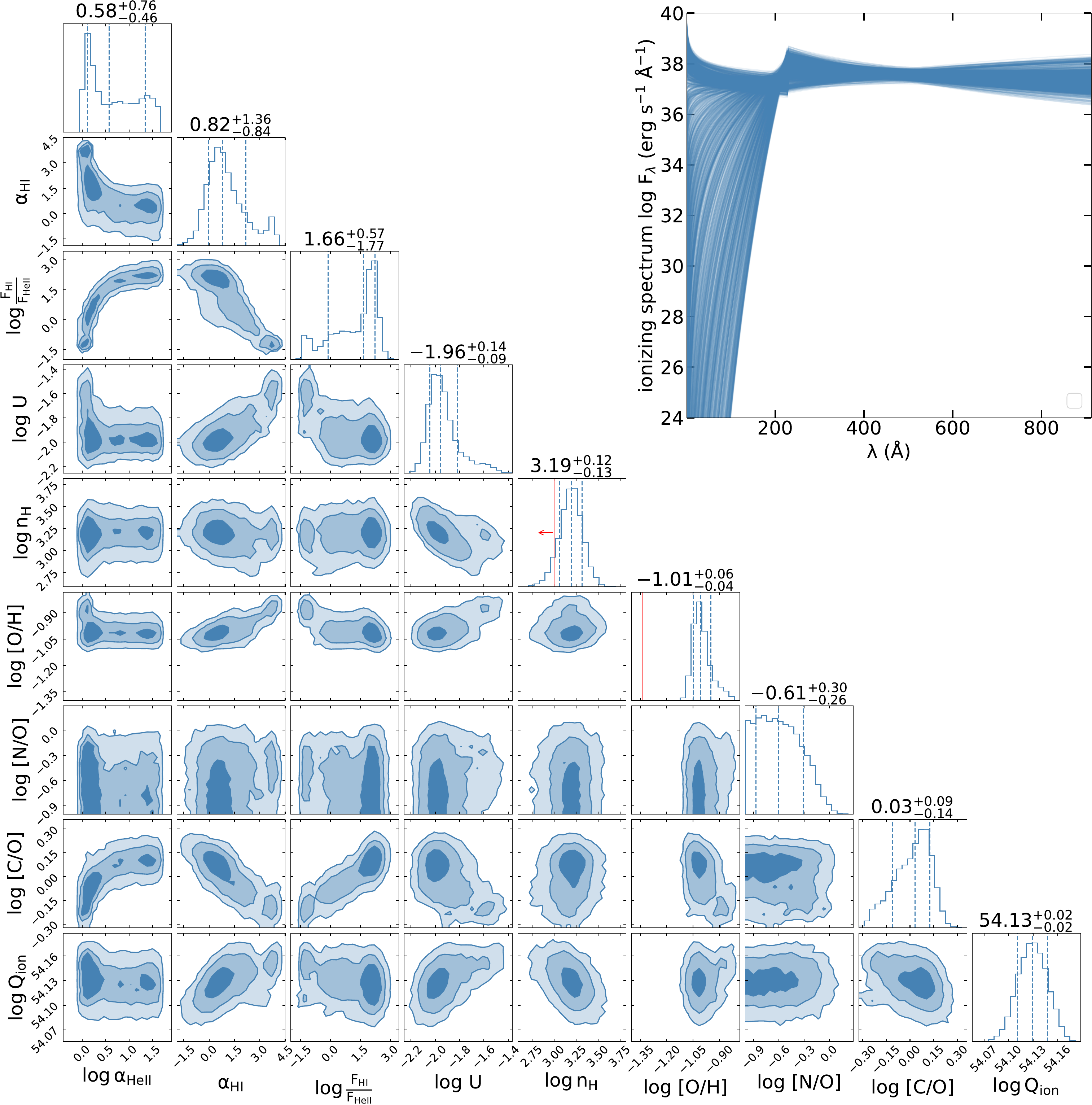}
    \caption{Posterior distribution of the \cue{} fit parameters and the inferred ionizing spectrum. The contours show the 1, 2, and 3$\sigma$ of the posterior. The blue dashed lines mark the 16\%/50\%/84\% quantiles. The red lines are the results for the same parameters from \citet{Cameron2024}, inferred in a different fashion using recipes applied to observed line ratios. The \cue\ fit is able to put tight constraints on $U$, $\nH$, [O/H], and $\Qion$. Since there is no nitrogen line detected, [N/O] shows a wide posterior. A solar [C/O] is preferred to explain the strong [CIV] lines. The upper right panel shows 2000 random draws of the ionizing spectrum posterior. Despite the large range of the ionizing spectrum parameters, the total ionizing budget of the two power laws is well constrained. The ionizing spectrum shows an AGN-like spectrum at $\lambda > \lambda_\mathrm{HeII}$ and a redder spectrum at $\lambda < \lambda_\mathrm{HeII}$.}
    \label{fig:cue_corner}
\end{figure*}

The full posterior parameter distribution is shown in Figure~\ref{fig:cue_corner}. \cue\ is able to put meaningful constraints on all of the parameters except for the slope of the HeII ionizing spectrum. The fit infers a high ionization parameter $\log U \approx -2$, consistent with the typically high degrees of ionization for high-redshift galaxies (e.g., \citealt{Sanders2023}). The gas-phase density $\log \nH = 3.19$\,cm$^{-3}$ and the low gas-phase metallicity $\log [\mathrm{O}/\mathrm{H}] \approx -1$ generally agree with \citet{Cameron2024}. The low [N/O] is expected given the nondetection of nitrogen lines, and a nitrogen deficit is also consistent with expectations from the secondary production of nitrogen (e.g., \citealt{Nicholls2017}). However, some galaxies at higher redshifts ($z>6$) show super-solar N/O (e.g., \citealt{Bunker2023a}; \citealt{Topping2024}; \citealt{Castellano2024}), suggesting there may be a wide range of potential nitrogen enrichment histories. Unlike the sub-solar [C/O] value \citet{Cameron2024} inferred from the CIII] 1909\,\AA / [OIII] 5007\,\AA, we estimate a solar [C/O]. Likely this difference is driven by the additional inclusion of the strong [CIV] lines in our fits, forcing the abundance of carbon to increase to explain the observed fluxes. Meanwhile, the CIII] doublets and [OIII] lines remain well fit by \cue\ as shown in figure~\ref{fig:cue_lines}.

\subsection{Inferred Ionizing Spectra}\label{subsec:cue_ionspec}
Figure~\ref{fig:cue_corner} also shows the posterior of our ionizing spectrum translated from the parameters $\alpha_\mathrm{HeII}$, $\alpha_\mathrm{HI}$, $F_\mathrm{HI}/F_\mathrm{HeII}$, and $\Qion$. We inferred a flat ionizing spectrum at $228\,\mathrm{\AA}<\lambda<912\,\mathrm{\AA}$, consistent with ionizing sources that can produce hard photons like AGNs, X-ray binaries, and metal-poor stars. 
Despite the wide posterior of $\alpha_\mathrm{HeII}$ and $F_\mathrm{HI}/F_\mathrm{HeII}$, the predicted production rate of hard ionizing photons $\log Q_\mathrm{HeII} = 52.62^{+0.07}_{-0.08}$\,s$^{-1}$ and soft ionizing photons $\log Q_\mathrm{HI} = 54.12^{+0.02}_{-0.02}$\,s$^{-1}$\footnote{Here $\QHeII$ and $Q_\mathrm{HI}$ are defined following the same wavelength range of $F_\mathrm{HeII}$ and $F_\mathrm{HI}$.}
tightly constrains the basic shape of the ionizing spectrum. 

The high rate of hard ionizing photons can be explained by an AGN. AGN ionizing spectra are conventionally assumed to have a spectral index of $1.2 \le \alpha_\mathrm{AGN} \le 2$ in $F_\nu$ (e.g., \citealt{Groves2004}; \citealt{Feltre2016}). Our inferred ionizing spectrum shows that the ionizing source of GS9422 is marginally consistent with an AGN ionizing spectrum at $\lambda > \lambda_\mathrm{HeII}$ but redder at $\lambda < \lambda_\mathrm{HeII}$. It thus requires a composite ionizing spectrum including also ionizing photons from a stellar population, explored in the following section.

To put these results in context, we additionally compare the \cue\ predictions with the emission lines powered by an SSP ($t_\mathrm{age} = 3.2$\,Myr and $\log (Z / Z_\odot) = -1.34$) of top-heavy IMF as approximately inferred by \citet{Cameron2024}. The SSP is generated from the MIST isochrones \citep{Choi2016} and the C3K stellar spectral libraries with the Flexible Stellar Population Synthesis (FSPS; \citealt{Conroy2009, Conroy&Gunn2010}) framework. We use this set of isochrones and stellar libraries throughout this paper. We adopt an IMF form of a single power-law over the entire mass range $\frac{\mathrm{d}N}{\mathrm{d}m} \propto (\frac{m}{\mathrm{M}_\odot})^{\delta}$, with a lower mass limit $M_l = 0.08$\,$\mathrm{M}_\odot$, and an upper mass limit of $M_u =300$\,$\mathrm{M}_\odot$ following \citet{Cameron2024}. A high $M_u$ generally leads to more HeII ionizing photons as it allows the presence of very luminous giants and WR stars \citep{Eldridge2017}, and is supported by the identification of very massive stars in intense starburst regions \citep{Crowther2016}. According to the IMF excess and SSP properties \citet{Cameron2024} estimated, we generate an SSP at $t_\mathrm{age} = 3.2$\,Myr and $\log (Z / Z_\odot) = -1.34$ assuming an extraordinary top-heavy IMF $\delta \approx 1.5$. The slope is derived according to the expectation from \citet{Cameron2024} that the top-heavy SSP has 112 $\mathrm{M}_\odot$ per hot massive star ($> 100$\,$\mathrm{M}_\odot$). Remarkably, this slope of 1.5 deviates significantly from the Salpeter IMF index of -2.35, where a slope of approximately -2.3 is the high-mass slope used in all standard extragalactic IMFs (e.g, \citealt{Salpeter1955}; \citealt{Kroupa2001}; \citealt{Chabrier2003}). Under this assumption, 99.8\% of the mass in this IMF is contained in stars of $> 50$\,M$_\odot$.

Using the ionizing spectrum from this top-heavy SSP along with the median of \cue\ nebular parameter posterior, \cloudy\ computes emission lines that match reasonably well with the observations. This top-heavy SSP produces more HeII ionizing photons and has a redder slope at $\lambda>\lambda_\mathrm{HeII}$ compared to the \cue\ prediction as shown in Figure~\ref{fig:ion_spec}, leading to higher [CIV] and HeII 1640\,\AA\ fluxes of the top-heavy scenario in Figure~\ref{fig:cue_lines}. 

\citet{Cameron2024} argues that the shape of the UV continuum suggests that it is dominated ($>$80\%) by nebular continuum. However, even with this extreme IMF we are only able to get $\sim$70\% contribution, and furthermore are underpredicting the observed UV flux by a factor of two. Even though it does not entirely satisfy the observational constraints and the nebular condition is not optimized to match the emission line measurements, we illustrate the top-heavy IMF solution in Figure~\ref{fig:cue_lines} for comparison purposes. This issue could be mitigated by using BPASS spectral synthesis \citep{Eldridge2017} as the BPASS SSP at this age and metallicity produces a stronger UV emission at the fixed $\QH$. However, we caution that there is significant uncertainty in the ionizing and UV fluxes of massive metal-poor stars (e.g., \citealt{Umeda2022}; \citealt{Olivier2022}; \citealt{Drout2023}). In the next section, we propose a solution that does not require $>80\%$ dominance by nebular emission.

\section{Decomposing the implied ionizing and UV radiation into stellar and AGN components}\label{sec:ssp_agn}
In the previous section it was demonstrated that a parametric form of the ionizing spectrum well-predicts the emission line fluxes. Here we interpret this ionizing spectrum by decomposing it into stellar and AGN light. 

The simplest model has the ionizing source(s) reproducing both the ionizing continuum and the observed UV emission, for example, $F_\nu$ at 1500\,\AA, i.e., $\Fuv$. Given the moderate $\rHeII$ from the \cue\ fit and high ionizing photon production efficiency $\eff \equiv \Qion/\Fuv$ suggested by \citet{Cameron2024}, we consider a mixed contribution by stars and an AGN to the ionizing radiation. This scenario is common and often found in other galaxies (e.g., \citealt{Brinchmann2023}; \citealt{Duncan2023}) and does not require an exotic system with a top-heavy IMF. In this section, we fit the ionizing spectrum and $\eff$ based on \cue\ results with a combination of SSPs and AGNs, and investigate the relative fractions of stellar and AGN light needed to explain GS9422.

\subsection{Separating AGN and stellar emission}\label{subsec:ssp_agn_fit}
Here we aim to find a combination of the SSP and AGN that describes both the observed emission lines and observed UV radiation of GS9422. We note that the goal is not to find all plausible combinations of stellar and AGN light, but instead to produce a single explanation that is consistent with the observed data and does not require IMF variations.

    \label{fig:QHeII}

We generate SSPs assuming a \citet{Kroupa2001} IMF, based on the MIST isochrones \citep{Choi2016} and the C3K stellar spectral libraries. Because young stars dominate the ionizing photon production in stellar populations and GS9422 has a low gas-phase metallicity, we consider all SSPs with ages younger than 10\,Myr and stellar metallicities $-2 \le \log Z / Z_\odot < -0.4$. These are normal young stellar populations without IMF variations.
The AGN emission from the accretion disk is approximated by a power-law $F_{\nu, \mathrm{AGN}} = A_\mathrm{AGN} \lambda^{\alpha_\mathrm{AGN}}$, with the power-law index $1.2 \le \alpha_\mathrm{AGN} \le 2$ at $1 \le \lambda \le 2500$\,\AA\ (e.g., \citealt{Groves2004}; \citealt{Stevans2014}; \citealt{Feltre2016}). We then combine the SSP and AGN spectrum to generate the composite ionizing and UV spectrum, 
\begin{equation}
    \frac{F_\nu}{\QH} = \fstar \frac{F_{\nu, \mathrm{SSP}}}{Q_\mathrm{H, SSP}} + (1-\fstar) \frac{F_{\nu, \mathrm{AGN}}}{Q_\mathrm{H, AGN}},
\end{equation} 
where $\fstar = \frac{Q_\mathrm{H, SSP}}{\QH}$ defines the stellar fraction normalized by the number of ionizing photons. 

The PRISM spectrum has $\Fuv = 3.24 \times 10^{28}$\,erg s$^{-1}$ Hz$^{-1}$ \citep{Cameron2024}. We allow an uncertainty of 10\% in the $\Fuv$ spectral flux density motivated by uncertainties in slit losses and flux calibrations in the literature (e.g., \citealt{Bunker2023}; \citealt{Curtis-Lake2023}; \citealt{deGraaff2024}). Then we sample $\Qion$ from the \cue\ posteriors and calculate the distribution of $\eff$ as our fit data.

Since the \cue\ prediction has a large uncertainty in the slope of the HeII ionizing spectrum but with a relatively small uncertainty in $\QHeII$, we set a constraint on the HeII photon ratio $\rHeII$ of the combined SSP and AGN fit. This requirement is important for predicting the correct HeII line fluxes with the SSP and AGN decomposition.
In summary, the fit data includes
\begin{equation}\label{eq:}
    \boldsymbol{Y} = \{\frac{F_{\nu, \mathrm{ionizing}}}{\QH}, N \log \rHeII, N \log \eff\},
\end{equation}
where $N$ is the number of the spectra flux points of $F_{\nu, \mathrm{ionizing}}$. 

We run a series of fits for each SSP in which we attempt to reproduce the \cue\ posterior median ionizing spectrum shape (see Section~\ref{sec:cue_res}), the HeII ionizing photon contribution to $\QH$, and $\eff$. 
The free parameters of the fit are the stellar contribution to the ionizing photons $\fstar$ and the AGN spectrum index $\alpha_{\mathrm{AGN}}$, which are allowed to vary over $0 \le \fstar \le 1$ and $1.2 \le \alpha_\mathrm{AGN} \le 2$. 
We assume $\boldsymbol{Y}$ follows a Gaussian likelihood:
\begin{equation}
    \ln \mathcal{L} = -\frac{\ln (2 \pi \sigma_\mathrm{\boldsymbol{Y}, Cue}^2)}{2} - \frac{ \left(\boldsymbol{Y}_{\mathrm{SSP+AGN}} - \boldsymbol{Y}_{\mathrm{Cue}}\right)^2}{2 \sigma_\mathrm{\boldsymbol{Y}, Cue}^2}
\end{equation}
As we have mentioned, $\boldsymbol{Y}_\mathrm{\cue}$ is derived from the observations and the maximum likelihood estimation from \cue\ in Section~\ref{sec:cue_fit}. The uncertainty $\sigma_\mathrm{\boldsymbol{Y}, Cue}$ is the 1$\sigma$ range of the posterior of $\boldsymbol{Y}_\mathrm{\cue}$. The fit $\boldsymbol{Y}_\mathrm{SSP+AGN}$ is calculated based on the sum of the normalized SSP spectrum and AGN spectrum.

\begin{figure*}
    \centering
    \includegraphics[width=0.9\textwidth]{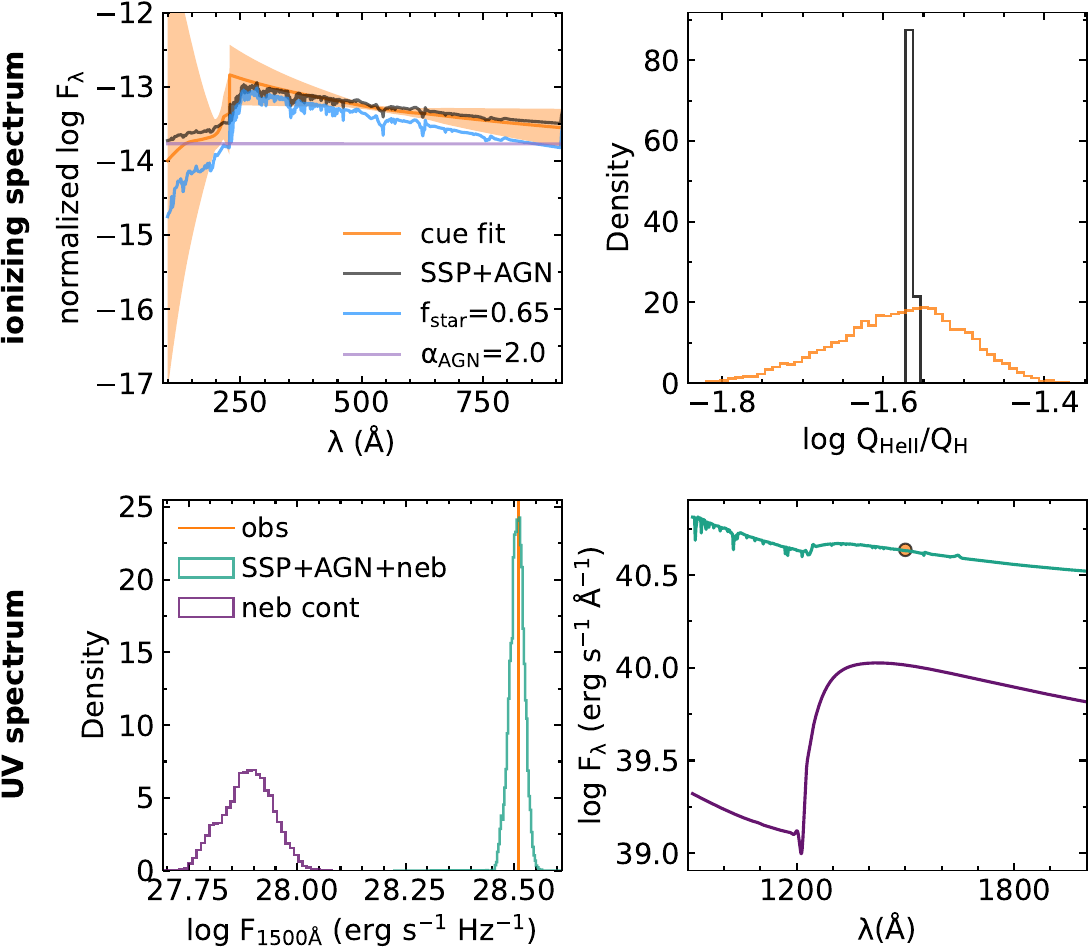}
    \caption{The nebular and UV emission of GS9422 can be powered by a combination of stars and AGN. Here we find the inferred ionizing spectrum from \cue{} and the observed $\Fuv$ can both be reproduced by the sum of a $t_\mathrm{age} = 3.55$\,Myr, $\log Z/Z_\odot = -1.97$ SSP and an AGN with a spectral index $\alpha_\mathrm{AGN} = 2$. Top left: the ionizing spectrum shape of the \cue\ posterior median (orange), SSP+AGN (black), SSP (blue), and AGN (purple). The shaded area reflects the 1$\sigma$ uncertainty of the \cue\ fit. The SSP produces $\sim$65\% of the ionizing photons and AGN contributes the rest 36\%. Top right: the $\rHeII$ distribution of the \cue\ posterior (orange) and SSP+AGN fit results (black). The \cue\ posterior distribution is rescaled for presentation. Bottom left: the UV continuum constraints. The measurement is in orange. The sum of the SSP, AGN, and nebular continuum is in green. The nebular emission contributes $\sim$24\% of the observed UV continuum as shown in purple. Bottom right: the predicted total UV continuum (green) and sub-contribution of the nebular continuum (purple).}
    \label{fig:ssp_AGN}
\end{figure*}

After sampling the posterior of $\fstar$ and the $\alpha_\mathrm{AGN}$ for each SSP with \texttt{dynesty}, we select the best fit from the Bayesian evidence. The fit with higher evidence is considered more plausible given the data. We show the results for the fit associated with this best-fit SSP in Figure~\ref{fig:ssp_AGN}.  

\subsection{A mixture of stars and AGN can power the observed emission in GS9422}
Figure~\ref{fig:ssp_AGN} shows the median ionizing spectrum from the \cue\ fit in Section~\ref{sec:cue_fit} decomposed into the best-fitting combination of light from a simple stellar population and AGN. It turns out that the $\rHeII$ has a considerable amount of constraining power on the results. The best-fitting SSP according to the Bayesian evidence has $t_\mathrm{age} = 3.55$\,Myr and $\log Z/Z_\odot = -1.97$. The ionizing spectrum from this SSP is shown in Figure~\ref{fig:ssp_AGN}. Our fit infers $\fstar \sim 0.65$ and $\alpha_\mathrm{AGN} \sim 2$, i.e. the light is mostly stellar with a sub-dominant contribution from a fairly red AGN. To put the inferred $\fstar$ into context, we derive the stellar mass of the decomposed SSP to be $M_* = 10^{7.2}$\,$\mathrm M_\odot$ by scaling the SSP to match the normalization of the stellar component of the inferred ionizing continuum. The estimated $M_*$ is below the total stellar mass of GS9422 from SED fittings (\citealt{Scholtz2023}; \citealt{Simmonds2024}; \citealt{Terp2024}). This is expected because the stellar component of the ionizing spectrum is dominated by young O and B stars, while the actual star formation history of the galaxy is much extended than the SSP and contains older stellar populations that contribute to the total stellar mass. Similarly, we use AGN SED templates to derive the bolometric luminosity of the decomposed AGN $L_\mathrm{bol}$. We scale the AGN SEDs with $\alpha_\mathrm{AGN} > 1.7$ from \citet{Elvis1994} and \citet{Richards2006} to match with the normalization of the inferred AGN ionizing continuum. This leads to $L_\mathrm{bol} \approx (1 \pm 0.6) \times 10^{44}$\,erg/s, consistent with the low $L_\mathrm{bol}$ of Type-2 AGNs (e.g. \citealt{Lusso2012}). The estimated $L_\mathrm{bol}$ translates to a lower limit on the black hole mass $M_\mathrm{BH} \gtrsim 10^6$\,$M_\odot$ if this black hole is radiating below the Eddington limit.

In order to ensure this decomposition remains consistent with the observed emission lines and UV flux density, we run \cloudy\ directly by inputting this SSP+AGN ionizing spectrum and the nebular properties from the \cue\ posterior median. The output emission lines are compared with the observation and \cue\ fit in Figure~\ref{fig:cue_lines}. This combination of stars and AGN can explain the emission lines with a reduced $\chi^2$ comparable to the \cue\ fits (Section~\ref{sec:cue_res}). The majority of the lines agree with the observations within $1\sigma_\mathrm{obs}$. By attributing 65\% of the ionizing photons to stars, we avoid the problems of a pure AGN model, such as the over-prediction of the hard ionizing photons, and high [OII] and [SII] fluxes associated with a low $U$.

Despite the relatively good fit, several high ionization state lines, including the [CIV] lines and HeII 4868\,\AA\ remain underestimated by $\gtrsim 3\sigma_\mathrm{obs}$. This is because even though the SSP+AGN decomposition reproduces the $\rHeII$, the slope of the HeII ionizing spectrum is slightly softer than the \cue\ inferred $\alpha_\mathrm{HeII}$. These high-ionization state lines are sensitive to the temperature structure, which is affected by the ionizing radiation shape via ionizing radiation deposition across the gas. However, we do not consider this to be a significant drawback; as discussed in Section~\ref{subsec:ssp_agn_fit} there remain significant uncertainties in the hard ionizing outputs of low-metallicity stellar populations and this deviation is well within those uncertainties. This is additionally mitigated by the fact that the stellar ionizing photons in a realistic source will come from composite stellar populations (CSPs) rather than SSPs. In this way our model combining a mixture of AGN and SSP emission primarily serves as a plausible example to demonstrate the efficacy of this mixture model for ionizing sources without requiring an exotic IMF.

Figure~\ref{fig:ssp_AGN} also shows the predicted UV continuum resulting from the best-fit combination of stars and AGN. Our fit is able to well reproduce $\Fuv$ from the PRISM spectrum by combining three components: the UV emission from the stars, the UV emission from the AGN, and the nebular continuum predicted using the \cue\ parameters. The inferred $F_\mathrm{1500, \mathrm{AGN}}$ is consistent with a low-luminosity AGN. 

We further compare the observed $\Fuv$ with the posterior of the nebular continuum at 1500\,\AA\ inferred by \cue\ (Section~\ref{sec:cue_res}). In contrast to the model presented in \citet{Cameron2024} which has $>80$\% nebular contribution in the UV, \cue\ predicts the nebular continuum at 1500\AA\ to contribute $\sim$24\% of the observed value. This is an important point as the dominance of the UV by two-photon emission is a key part of the argument for a top-heavy IMF. We present an alternative explanation for the mild UV turnover prediction at $\lambda \approx 1430$\,\AA\ in the discussion.  

\section{Conclusion and Discussion}\label{sec:summary}
In this paper, we investigate the nature of the ionizing spectrum of GS9422 ($z = 5.943$), which has previously been interpreted as an exceptionally top-heavy IMF \citep{Cameron2024}. We model the emission line fluxes of GS9422 with \cue\ v0.1, a flexible neural net-powered CLOUDY photoionization model. This model covers a large range of the ionization parameter, gas density, gas-phase metallicity, N/O, and C/O, and importantly, is agnostic to the source of the ionizing spectrum, characterizing the ionizing spectrum with a four-part piece-wise power-law. This allows \cue\ to perform an exploration of the full parameter space of ionizing sources. We modify \cue\ to allow only a double power-law ionizing spectrum which is sufficient to describe this source, finding $\rHeII = -1.51$, a high ionization parameter $\log U = -1.96$, a gas density $\log \nH = 3.19$\,cm$^{-3}$, $\log [\mathrm{O/H}] = -1.01$,  $\log [\mathrm{N/O}] = -0.61$, $\log [\mathrm{C/O}] = 0.03$. \cue\ is able to achieve a good fit the the observed lines and UV continuum at 1500 $\AA$ using an ionizing spectrum which we can decompose in a straightforward fashion into a combination of low-metallicity young stellar population and an AGN.

\subsection{Interpreting GS9422 with a mixture of stars and AGN with DLA}\label{subsec:DLA}
We show that an approximately 3:2 fraction of the stellar to AGN contribution to the ionizing photon budget reproduces both the emission line fluxes and $\Fuv$. It should be emphasized that the exact stellar contribution needed to explain this galaxy depends on the stellar models. However, models of ionizing emission from stars, especially massive stars, is highly uncertain (e.g., \citealt{Eldridge2017}; \citealt{EldridgeStanway2022}; \citealt{Wheeler2023}). The total number and distribution of their ionizing photons is highly sensitive to the details of stellar evolution, and affected by several complex factors which are difficult to model such as mass loss, rotation, and multiplicity (e.g., \citealt{Smith2014}). As an example, stellar rotation which was recently included in MIST isochrones can prolong ionizing photon production and produce a harder ionizing spectrum (e.g., \citealt{Leitherer2014}; \citealt{Choi2017}).

\begin{figure}
    \centering
    \includegraphics[width=0.85\columnwidth]{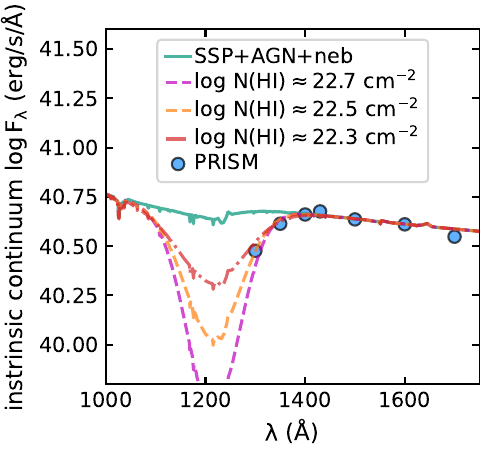}
    \caption{The observed UV turnover around $\lambda_\mathrm{rest} \approx 1430$\,\AA\ is not consistent with intrinsic stellar or AGN light. However, it can be reproduced by a damped Lyman-$\alpha$ absorber near the ionizing source. PRISM data \citep{Cameron2024} in the rest-frame is in blue. The UV radiation of the composite stellar and AGN solution and the nebular continuum are shown in green. Note that we do not show emission lines here. The UV continuum after being absorbed by a DLA of different $N$(HI) is in magenta, orange, and red. We estimate an absorber column density of $\log N(\mathrm{HI}) \lesssim 22.7$\,cm$^{-2}$. We do not model the Gunn-Peterson trough due to absorption from neutral hydrogen in the IGM, which might further lower the DLA $N$(HI) requirement. In a DLA of this relatively high column depth, it would be unusual but not impossible to have no dust. Further discussion of this point is in Section~\ref{subsec:DLA}. }
    \label{fig:DLA}
\end{figure}

In our combined SSP and AGN model the observed nebular continuum is only 24\% of the UV flux. The inferred UV continuum shows a mild UV turnover (see Figure~\ref{fig:ssp_AGN} and \ref{fig:DLA}) attributed to the two-photon continuum peaked at $\lambda_\mathrm{rest} \approx 1430$\,\AA\ (e.g., \citealt{Guzman2017}, \citealt{Byler2017}). As \citet{Cameron2024} showed, $\gtrsim80\%$ nebular contribution to the UV continuum is required for the two-photon continuum to display the steep UV turnover observed in the PRISM spectrum. In contrast, stellar and AGN emission are both expected to show pure power-law behavior at these wavelengths with no turnover.

In Figure~\ref{fig:DLA}, we show the PRISM spectrum near the UV turnover and the total UV continuum from the stellar population, AGN, and nebular continuum in rest-frame. Our predicted UV flux agrees with the PRISM spectrum around $\lambda_\mathrm{rest} = 1430$\,\AA\ but not bluewards of this wavelength. One way to add a turnover to otherwise pure power-law stellar and AGN spectra is a damped Lyman-$\alpha$ profile blueward of 1430\,\AA\ will allow us to reproduce the UV turnover, such as resonant scattering and absorption by the IGM or damped Lyman-$\alpha$ absorber (DLA) (e.g., \citealt{Heintz2024}). As \citet{Cameron2024} suggested, neutral IGM at $z=5.943$ alone is not able to capture the broad width of this absorption. Therefore, we consider including DLA to explain the UV turnover. 

We model DLAs with a range of different neutral hydrogen column densities $N$(HI) and compare them in Figure~\ref{fig:DLA}. While \citet{Cameron2024} finds that a very high $\log N(\mathrm{HI}) \gtrsim 23$\,cm$^{-2}$ and perhaps unphysical column density is needed, here we find lower column densities are sufficient, as our proposed solution has a bluer UV continuum slope comparing to the pure stellar ionizing source. Additionally, it is plausible that a partial contribution to the turnover from the nebular continuum will further lower the DLA column density requirement. DLAs of $\log N(\mathrm{HI}) > 22$\,cm$^{-2}$ have been found in other studies at $z \gtrsim 2$ (e.g., \citealt{Prochaska2009}; \citealt{Noterdaeme2012}; \citealt{Selsing2019}; \citealt{Heintz2024}; \citealt{Umeda2024}). DLA surveys have found Lyman-$\alpha$ emission from DLA host galaxies and some of these DLAs show small impact parameters (e.g., \citealt{Moller2004}; \citealt{Krogager2017}; \citealt{Oyarzun2024}). Given the required $N$(HI), and our inferred gas density $\sim 10^{3}$\,cm$^{-3}$ and low gas-phase metallicity, we would expect the DLA clouds in this galaxy to have a size of $\lesssim 100$\,pc and a low dust content. Note that our $N(\mathrm{HI})$ estimate is subject to the uncertainties in the UV spectrum of stellar models.

The ionizing sources and the gas are implicitly assumed to be spatially mixed in this model. Because there is little dust attenuation towards the observed emission lines, and because the DLA must be in front of the combined stellar and AGN continuum, the DLA also must have very little dust attenuation. This would be unusual but not impossible at these column densities. The relationship between DLA column depth and inferred dust attenuation is further discussed in \citet{Tacchella2024}, where it is estimated that this DLA likely has $A_\mathrm{V} < 1.25$ and can be as low as $A_\mathrm{V} = 0.1-0.5$. The \citet{Tacchella2024} scenario is described in detail in the next section.

In this analysis we focus on reproducing the emission lines and UV fluxes as this is where the core evidence for a top-heavy IMF has been presented. Our fiducial model including young stars and an AGN source does predict some basic properties of the optical spectrum, e.g. the observed Balmer jump, but alone is not sufficient to reproduce the optical observed spectrum. Instead, additional components are needed and indeed expected, such as emission from older stars. Also, the exact optical flux from our proposed scenario would depend on the AGN physics such as the temperature of the accretion disk. We defer this to future work; here we focus on the UV and ionizing spectrum as it is the main driver for the top-heavy IMF scenario.

Additionally, \cue\ predicts an intrinsic Lyman-$\alpha$ emission line flux $\approx 6 \times 10^{-17}$\,erg\,s$^{-1}$\,\,cm$^{-2}$, $\approx6$x brighter than the observation. The DLA column depth inferred from the continuum fit suggests that this intrinsic Ly-$\alpha$ flux should be attenuated further by a factor of 4, bringing it nearly in line with the observed Ly-$\alpha$ flux. However, we caution that Lyman-$\alpha$ is sensitive to significant resonance and scattering effects and a DLA absorber is one of several possibilities that could cause a significant reduction in observed Lyman-$\alpha$ flux (e.g., \citealt{Mason2018}).

\subsection{Other studies and future investigations}
In a companion study, \citet{Tacchella2024} leverage the spatially resolved JWST/NIRCam photometry and deep NIRSpec spectroscopy from JADES and JWST Extragalactic Medium-band Survey (JEMS; \citealt{Williams2023}) to investigate the nature of GS9422. They find morphological variations in different bands, suggesting spatial variation in the sources of the nebular, rest-UV, and rest-optical emission. Based on this, they propose that this galaxy is consistent with a normal galaxy hosting an AGN, where the AGN powers the off-planar nebular emission, the young stellar component dominates the rest-UV with the DLA clouds, and the older stellar component distributed in a disk shares the contribution to the rest-optical with the AGN. Their work agrees with the work presented here in explaining the UV and nebular emission with a mixture of young stellar population and AGN, though they propose a slightly redder AGN. 
Our work has a complementary focus on the detailed ionizing radiation and nebular properties of GS9422.

To be clear, current data alone cannot clearly refute either proposed scenario: a mixed stellar and AGN ionizing source, or a remarkably top-heavy (essentially top-only) IMF. An IMF as top-heavy as the one proposed in \citet{Cameron2024} would be an extraordinary discovery and, if such IMF variations were common, would change the inferred star formation rates from distant galaxies by large factors ($\gtrsim$5-10), requiring a full revision of our understanding of galaxy formation at early times. In this context it is very important to propose alternate solutions such as the one presented here and to continue testing their viability. Future higher resolution data would be helpful in measuring the ionizing properties of different locations across this galaxy to discriminate between these two scenarios, and additional infrared data can help to establish the presence of AGN. 
Finally, this work shows the power of \cue\ in interpreting complex sources powered by unusual or mixed ionizing spectra, and it is hoped that more such analyses will be useful to understand the peculiar nebular properties observed in high-redshift objects. 


\begin{acknowledgments}
We thank the anonymous referee for a helpful review. We thank Alex Cameron and Harley Katz for discussion and for clarifications about their work. Y.L. and J.L. are supported under Program number JWST-GO-01810.004-A provided through a grant from the STScI under NASA contract NAS503127. 
\end{acknowledgments}

\software{FSPS \citep{Conroy2009}, PYTHON-FSPS \citep{python-fsps}, dynesty \citep{Speagle2020,Koposov2022}, MIST \citep{Choi2016, Dotter2016}, Astropy \citep{astropy:2013, astropy:2018}, NumPy \citep{harris2020array}, SciPy \citep{2020SciPy-NMeth}, TensorFlow \citep{tensorflow2015-whitepaper}. \\~\\}

\bibliography{main}{}
\bibliographystyle{aasjournal}

\end{CJK*}
\end{document}